\documentclass[prc,twocolumn,floatfix,nofootinbib]{revtex4}
\usepackage{amsmath,amssymb,graphicx,bm}
\usepackage{color}
\usepackage{ulem}
\usepackage{graphicx}
\usepackage{epsfig}
\usepackage{epstopdf}

\DeclareMathOperator{\arccot}{arccot}
\newcommand{\adag}[1]{a^\dagger_{#1}}
\newcommand{\aop}[1]{a^{\vphantom{\dagger}}_{#1}}
\newcommand{\BMBPT}[1]{\text{BMBPT#1}}
\newcommand{\bdag}[1]{\beta^\dagger_{#1}}
\newcommand{\bop}[1]{\beta^{\vphantom{\dagger}}_{#1}}
\newcommand{\bra}[1]{\langle #1|}

\newcommand{\calE}{\mathcal{E}}
\newcommand{\calO}{\mathcal{O}}
\newcommand{\calV}{\mathcal{V}}
\newcommand{\Deltab}[1]{\Delta_{#1}}
\newcommand{\down}{\mathord{\downarrow}}
\newcommand{\Eb}[1]{E_{#1}}

\newcommand{\Eq}[1]{Eq.~(\ref{#1})}
\newcommand{\Eqs}[1]{Eqs.~(\ref{#1})}

\newcommand{\Fig}[1]{Fig.~\ref{#1}}

\newcommand{\HFB}{\text{HFB}}
\newcommand{\Hop}{\hat{H}}
\newcommand{\Kop}{\hat{K}}
\newcommand{\ket}[1]{|#1\rangle}
\newcommand{\kF}{k_{\mathrm{F}}}
\newcommand{\kv}{\vek{k}}
\newcommand{\Nop}{\hat{N}}
\newcommand{\pv}{\vek{p}}
\newcommand{\Qv}{\vek{Q}}
\newcommand{\qmax}{\Lambda'}

\newcommand{\qv}{\vek{q}}
\newcommand{\Sec}[1]{Sec.~\ref{#1}}
\newcommand{\Secs}[1]{Secs.~\ref{#1}}
\newcommand{\myRef}[1]{Ref.~\cite{#1}}
\newcommand{\Refs}[1]{Refs.~\cite{#1}}
\newcommand{\reff}{r_{\text{eff}}}
\newcommand{\UHF}[1]{U_{#1}}
\newcommand{\up}{\mathord{\uparrow}}
\newcommand{\vek}[1]{\bm{\mathrm{#1}}}
\newcommand{\vlowk}{{\ensuremath{V_{\text{low-}k}}}}
\newcommand{\Wop}{\hat{W}}
\newcommand{\xib}[1]{\xi_{#1}}

\begin{document}
\title{Low-momentum interactions for ultracold Fermi gases}
\author{M. Urban} \email{urban@ijclab.in2p3.fr}
\affiliation{Universit\'e Paris-Saclay, CNRS-IN2P3, IJCLab, 91405
  Orsay, France} \author{S. Ramanan} \email{suna@physics.iitm.ac.in}
\affiliation{Department of Physics, Indian Institute of Technology
  Madras, Chennai - 600036, India}
\begin{abstract} We consider a two-component Fermi gas with a
  contact interaction from the BCS regime to the unitary
  limit. Starting from the idea that many-body effects should not
  depend on short-distance or high-momentum physics which is encoded
  in the $s$-wave scattering length, but only on momentum scales of
  the order of the Fermi momentum, we build effective low-momentum
  interactions that reproduce the scattering phase shifts of the
  contact interaction below some momentum cutoff. Inspired from recent
  successes of this method in nuclear structure theory, we use these
  interactions to describe the equation of state of the Fermi gas in
  the framework of Hartree-Fock-Bogliubov theory with perturbative
  corrections. In the BCS regime, there is a range of cutoffs where we
  obtain fully converged results. Near unitarity, convergence is not
  yet reached, but we obtain promising results for the ground-state
  energies close to the experimental ones. Limitations and
  possible extensions of the approach are discussed.
\end{abstract}

\maketitle
\section{Introduction}
\label{sec:intro}
The study of ultracold Fermi gases opens new and exciting avenues into
the rich physics of fermions, where one can observe the effects of
quantum degeneracy and interactions and explore regions of strong
correlations, for example the crossover from the BCS to the BEC state,
including the unitary limit, where the scattering length
diverges~\cite{Strinati2018}. The progress made in the theoretical
understanding has been intimately coupled to the experimental
observation of these phenomena, using advanced trapping and cooling
techniques and the possibility to tune the interaction between alkali
atoms such as $^6$Li or $^{40}$K via Feshbach resonances.

In nuclear physics, it is well known that the $s$-wave scattering
length between two nucleons in the spin-singlet channel is very large
compared to the range of the interaction. Furthermore, in neutron
stars, the neutrons in the inner layers of the crust are in a strongly
correlated (almost) unitary regime. Therefore, much can be gleaned by
the connection between cold Fermi gases near the unitary limit and
low-energy nucleons.

Since in the case of ultracold atoms, the range of the interaction is
about four orders of magnitude smaller than the typical interparticle
spacing, the interaction between two fermions has been usually
modelled as a contact interaction with a coupling constant $g$. This
simplifies the many-body calculations, as the interactions get
restricted to the $s$ wave. The use of this so-called single-channel
model is valid in the case of a broad Feshbach
resonance~\cite{Simonucci2005}. However, such an interaction has to be
regularized, which can be done by choosing a momentum cutoff
$\Lambda$. Fixing the coupling constant for a given cutoff by the
requirement that it should reproduce the physical scattering length
$a$, one gets \cite{Pieri2000}
\begin{equation}
  \frac{1}{g} = \frac{m}{4\pi a} - \frac{m \Lambda}{2 \pi^2}\,,
  \label{eq:reg-contact}
\end{equation}
where $m$ is the atom mass. This shows that the coupling
constant vanishes when $\Lambda \rightarrow \infty$ and hence
particle-particle ladders have to be resummed in order to
get a non-vanishing contribution~\cite{Pieri2004}. However,
for a realistic description of the atom-atom interaction, the limit
$\Lambda\to\infty$ must be taken, since otherwise the finite cutoff
results in an effective range of the interaction, 
$\reff = 4/(\pi\Lambda)$.

In nuclear structure theory, the idea of using renormalization group
(RG) approaches to get low-momentum effective interactions has allowed
for major advances over the past two decades~\cite{Bogner2009}. In the
two-body sector, starting with the $s$-wave $T$-matrix equation,
\begin{multline}
  T_{0}(k,k';E) = V_{0}(k,k') \\
  + \frac{2}{\pi} \int_0^\Lambda\! dq  \, q^2
  \frac{V_{0}(k, q) T_{0}(q, k'; E)}{E-q^2/m}\,,
\end{multline}
where $V_0$ denotes the interaction in the $s$ wave, $E$ is the total
energy of the pair and $k$ and $k'$ are the incoming and outgoing
momenta in the center of mass frame, the intermediate states are cut
off at $\Lambda$. The requirement that the two-body observables (bound
states and phase shifts at momenta below the cutoff) must be
independent of the cutoff, leads to a $\Lambda$-dependent effective
low-momentum interaction called $\vlowk$. Therefore, contrary to what
is done in cold-atom physics, the cutoff $\Lambda$ for the
low-momentum interaction is not only finite, but typically lowered as
much as possible to include just the relevant momentum scales of the
problem. Such low-momentum interactions are ``soft'' and hence have
the advantage that many-body calculations become more perturbative.
  
To give an example, using RG softened interactions derived from chiral
perturbation theory, including the three-body force, one gets bound
nuclei already at the Hartree-Fock (HF) \cite{Coraggio2003} or
Hartree-Fock-Bogoliubov (HFB)%
  \footnote{In \myRef{Tichai2018}, the interaction was softened by the
    similarity renormalization group (SRG) instead of $\vlowk$.}
\cite{Tichai2018} level and obtains satisfactory results for
ground-state energies if one includes perturbatively corrections to
the HF(B) ground state. As another example, we mention our recent work
on screening of the pairing interaction in neutron matter
\cite{Ramanan2018}, where the use of a small cutoff allowed us to
retrieve the Gorkov-Melik-Barkhudarov result \cite{Gorkov1961} in the
low-density limit without the resummation of ladder diagrams in the
(3-particle 1-hole and 1-particle 3-hole) vertices.

The aim of this paper is to try this strategy, which is very
successful in nuclear physics, in the case of ultracold Fermi gases in
the BCS-BEC crossover. On the one hand, the situation is more
favorable in the case of cold atoms, namely in what concerns the
three-body force. While the $\Lambda$-dependence of $\vlowk$
completely accounts for the effects of the intermediate states beyond
$\Lambda$ in the two-body sector, this is not true in the many-body
sector. There, the RG running generates $\Lambda$ dependent three- and
higher-body forces \cite{Tolos2008,Hebeler2010,Hammer2013}. The
leading three-body term is of the form $(\psi^\dagger\psi)^3$, where
$\psi$ is the field operator. Because of the Pauli principle, this
term can only contribute if $\psi$ has at least three components,
which is the case in nuclear physics (neutrons and protons with spin
$\up$ and $\down$), but not in ultracold atoms with only two spin
states. On the other hand, the pairing correlations can become much
more important in ultracold atoms than in nuclei, especially near
unitarity and on the BEC side ($a>0$). Far on the BEC side, the lowest
excitations are molecules out of the condensate, which clearly require
a non-perturbative resummation of ladder diagrams. Therefore, we will
limit ourselves in the present work to the BCS side ($a<0$) of the
crossover, up to the unitary limit.

The paper is organized as follows. In \Sec{sec:sep-int}, we set up a
separable interaction which exactly reproduces the two-body scattering
phase shifts of a contact interaction up to the cutoff.  The elements
of HFB and the Bogoliubov many-body perturbation theory (BMBPT) are
set up in \Secs{sec:HFB} and~\ref{sec:bmbpt}. We present our results
in \Sec{sec:results} and our conclusions in \Sec{sec:concl}, where we
discuss also perspectives for future work. Some technical details and
lengthy equations are relegated to the appendices. Throughout the
paper, we use units with $\hbar=1$, where $\hbar$ is the reduced
Planck constant.
\section{Separable form of a contact interaction}
\label{sec:sep-int}
The scattering phase shifts of two particles 1 and 2 with opposite
spins $\up$ and $\down$, interacting via a contact interaction with
$s$-wave scattering length $a$, are given by
\begin{equation}
  \delta(q) = \arccot\Big(-\frac{1}{qa}\Big)\,,
  \label{phaseshiftcontact}
\end{equation}
where $q = q'$ is the momentum in the
center-of-mass frame, i.e., $\qv = (\pv_1-\pv_2)/2$ and $\qv' =
(\pv'_1-\pv'_2)/2$, if in- and outgoing momenta of the two particles
are denoted $\pv_{1,2}$ and $\pv'_{1,2}$, respectively.

We want to describe the system with a hamiltonian written in
second quantization as
\begin{multline}
  \Hop = \sum_{\pv\sigma} \frac{p^2}{2m} \adag{\sigma\pv}
  \aop{\sigma\pv}\\ + \sum_{\pv_1\pv_2\pv'_1\pv'_2}
  V_{\pv_1\pv_2\pv'_1\pv'_2} \adag{\pv_1\up} \adag{\pv_2\down}
  \aop{\pv'_2\down} \aop{\pv'_1\up}\,.
  \label{hamiltonian}
\end{multline}
This form is written for a finite volume $\calV$, but as usual, in the
limit of a large system, the summations over momenta will be
replaced by integrals:
\begin{equation}
  \sum_{\pv} \cdots \to \frac{\calV}{(2\pi)^3} \int\! d^3p\, \cdots\,.
  \label{summegral}
\end{equation}
Since a contact interaction acts only in the $s$ wave ($l=0$), we
write it in the conventions that are common if one works in a
partial-wave basis as
\begin{equation}
  V_{\pv_1\pv_2\pv'_1\pv'_2} = \frac{1}{\calV}\, 4\pi V_0(q,q')\,\delta_{\Qv,\Qv'}\,,
  \label{Vpartialwaves}
\end{equation}
where $\Qv = \pv_1+\pv_2$ and $\Qv' = \pv'_1+\pv'_2$ are the
incoming and outgoing total momenta. The factor $1/\calV$ 
in \Eq{Vpartialwaves}
ensures that $V_0(q,q')$ (having dimension energy times volume) is
independent of $\calV$.

Our aim is to construct a separable interaction of the form
\begin{equation}
  V_0(q,q') = g_0 F(q) F(q')\,,
  \label{Vseparable}
\end{equation}
which reproduces the phase shifts (\ref{phaseshiftcontact}) below some
cutoff $\Lambda$, but tends towards zero above this cutoff. In
\Eq{Vseparable}, $g_0$ denotes the coupling constant and $F(q)$ the
form factor (normalized to $F(0)=1$). If the interaction tends to zero
at high momenta, this implies that the phase shifts also tend to
zero.

\begin{figure}
  \epsfig{width=7.8cm,file=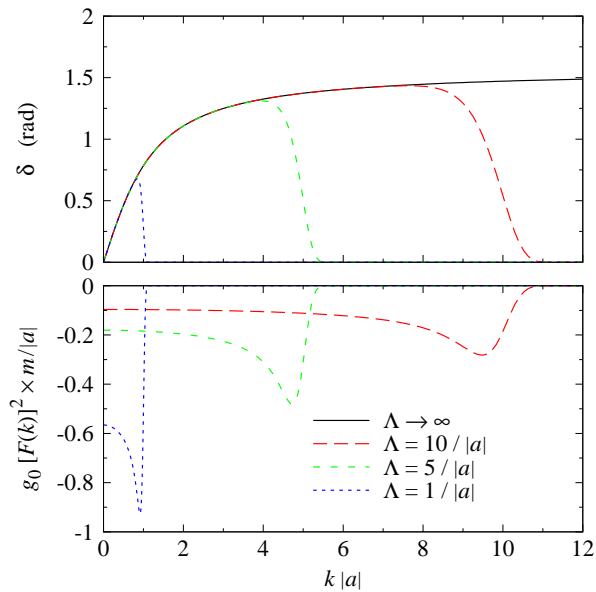}
  \caption{\label{fig:delta-F} Momentum dependence of the phase shifts
    multiplied by the regulator, and of the corresponding diagonal
    matrix elements of the potential, for different cutoffs
    (quantities made dimensionless by multiplication with the
    appropriate powers of $|a|$ and $m$).}
\end{figure}

Rather than determining $F(q)$ from the RG evolution as it is
  done for $\vlowk$, we find it easier to impose the behavior of the
phase shifts, namely as follows:
\begin{equation}
  \delta(q) = R\Big(\frac{q}{\Lambda}\Big)
    \arccot\Big(-\frac{1}{qa}\Big)\,,
\end{equation}
where $R(x)$ is a regulator function. This approach was also followed
in \Refs{RuizArriola2017,Koehler2007,Koehler2008,Koehler2010}, where
the authors used a sharp regulator $\theta(1-x)$. However, for better
numerical convergence near $q = \Lambda$, we prefer an exponential
regulator of the form $R(x) = \exp(-x^{2n})$, where $n$ is a parameter
that determines how smoothly $R(x)$ drops from 1 to 0 near $x=1$. The
phase shifts of a contact interaction (black solid line) and examples
of phase shifts cut off at different $\Lambda$ (dashed lines) are
displayed in the upper panel of \Fig{fig:delta-F} for the case
$n=10$. One sees that in practice the regulator can be set to zero
beyond some momentum $\qmax$, where the ratio $\qmax/\Lambda > 1$
depends on the smoothness parameter $n$. In our calculations, we
  choose $\qmax$ such that $R(\qmax/\Lambda) = 10^{-10}$, which in the
  case $n=10$ gives $\qmax \approx 1.17\,\Lambda$.

The problem of finding the separable interaction corresponding to
given phase shifts $\delta(q)$ was solved long ago by Tabakin
\cite{Tabakin1969}: the diagonal elements of the interaction, for
particles of mass $m$, can be computed with a principal-value integral
as
\begin{equation}
   V_0(q,q) = -\frac{\sin\delta(q)}{m q} \exp\left(\frac{2}{\pi}\,
   \mathcal{P}\!\!\int_0^\infty\! dq'\, \frac{q'
     \delta(q')}{q^2-q^{\prime\,2}}\right)\,,
   \label{inversescattering}
\end{equation}
from which follow the coupling constant and form factor
\begin{equation}
  g_0 = V_0(0,0)\,,\quad F(q) = \sqrt{V_0(q,q)/g_0}\,.
  \label{g0F}
\end{equation}
In the case $a>0$, when the potential has a bound state with binding
energy $1/(ma^2)$, the right-hand side of \Eq{inversescattering} gets
an additional factor $1+1/(qa)^2$ \cite{Tabakin1969,Bogner2006}.
Equation (\ref{inversescattering}) was used in \Refs{Koehler2007}
  and \cite{RuizArriola2017} to derive an analytical expression for
the separable interaction in the unitary limit with a sharp
cutoff.

As an illustration, we display in the lower panel of \Fig{fig:delta-F}
the diagonal matrix elements of the potential corresponding to the
phase shifts with different cutoffs shown in the upper panel. The
coupling constant increases when the cutoff is lowered, thereby
compensating the missing contribution from intermediate states which
are cut off by the regulator. Nevertheless, the procedure explained
here is not equivalent to the simpler prescription
\eqref{eq:reg-contact}, corresponding to $g_0 = g/(4\pi)$ and $F(q) =
\theta(\Lambda-q)$, preserving only the scattering length $a$ but not
the momentum dependence of the phase shifts up to the cutoff. In the
present case, the form factor $F(q)$ is a non-trivial function of
momentum, which ensures that not only the scattering length $a$, but
the entire momentum dependence of the phase shifts remains cutoff
independent, as it is the case with $\vlowk$. Actually, $\vlowk$ for
the neutron-neutron interaction in $s$ wave resembles very much our
separable interaction if $\Lambda\ll 1/\reff$, as shown in Fig.~11 of
\myRef{Ramanan2018}.

\section{Hartree-Fock-Bogoliubov}
\label{sec:HFB}
It is well known that BCS mean-field theory can qualitatively describe
the BCS-BEC crossover. In the cold-atom literature, this theory is
usually defined by the gap and number equations
[\Eqs{gap-equation-renormalized} and \eqref{number-equation} below]
with a contact interaction regularized according to
\Eq{eq:reg-contact} in the limit $\Lambda\to\infty$. However, in the
weak-coupling regime (i.e., $1/(\kF a)\ll -1$, where $\kF$ is the
Fermi momentum), the gap is exponentially suppressed and the dominant
energy correction compared to the ideal gas comes from the normal part
of the self-energy (i.e., the diagonal part in Nambu-Gorkov
formalism). In the limit $\Lambda\to\infty$, the calculation of this
self-energy requires the resummation of ladder diagrams as done, e.g.,
in \myRef{Pieri2004}, whereas in the present framework, we obtain it
already at the HF level. The mean-field theory including both the HF
self-energy and the pairing gap is called HFB theory.

Following \cite{FetterWalecka}, we start by defining the quasiparticle
operators%
\footnote{In the notation of \myRef{FetterWalecka}, our operators
  $\bop{\kv\up}$ and $\bop{\kv\down}$ correspond to the operators
  $\alpha_{\kv}$ and $\beta_{\kv}$, respectively.}
\begin{equation}
  \bop{\kv\up} = u_k \aop{\kv\up}-v_k\adag{-\kv\down}\,,\quad
  \bop{\kv\down} = u_k \aop{\kv\down}+v_k\adag{-\kv\up}\\
\end{equation}
with $u_k^2+v_k^2 = 1$. This Bogoliubov transformation can be inverted
to rewrite the creation and annihilation operators $\adag{\kv\sigma}$
and $\aop{\kv\sigma}$ in terms of the quasiparticle creation and
annihilation operators $\bdag{\kv\sigma}$ and $\bop{\kv\sigma}$. For
example, the annihilation operators can be expressed as
\begin{equation}
  \aop{\kv\up} = u_k \bop{\kv\up}+v_k\bdag{-\kv\down}\,,\quad
  \aop{\kv\down} = u_k \bop{\kv\down}-v_k\bdag{-\kv\up}\,.
  \label{bogoliubovtrafo}
\end{equation}
Let us now consider the operator
\begin{equation}
  \Kop = \Hop-\mu \Nop
\end{equation}
with $\Hop$ the hamiltonian defined in \Eq{hamiltonian}, $\Nop$ the
particle-number operator
\begin{equation}
  \Nop = \sum_{\kv\sigma}
  \adag{\kv\sigma}\aop{\kv\sigma}\,,
\end{equation}
and $\mu$ the chemical potential. Following the usual procedure
\cite{FetterWalecka}, one obtains the expressions for the $u$ and $v$
factors:
\begin{equation}
u_k = \sqrt{\frac{1}{2}+\frac{\xib{k}}{2\Eb{k}}}\,,\quad
v_k = \sqrt{\frac{1}{2}-\frac{\xib{k}}{2\Eb{k}}}\,,
\label{uvfactors}
\end{equation}
where
\begin{equation}
  \xib{k} = \frac{k^2}{2m}+\UHF{k}-\mu\,,\quad
  \Eb{k} = \sqrt{\xib{k}^2+\Deltab{k}^2}\,,
  \label{qpenergy}
\end{equation}
with a HF-like mean field $\UHF{k}$ and the gap $\Delta_k$.

The gap equation reads
\begin{equation}
  \Deltab{k} = -\frac{2}{\pi}\int_0^{{\qmax}}\! dp\, p^2\,
    V_0(k,p)\frac{\Deltab{p}}{2\Eb{p}}\,.
\end{equation}
Because of the separable form of the interaction, it is evident that
the momentum dependence of the gap $\Deltab{k}$ is given by
\begin{equation}
\Deltab{k} = \Deltab{0} F(k)\,,
\end{equation}
and solving the gap equation amounts to simply finding the number
$\Deltab{0}$ for which the following equation is satisfied:
\begin{equation}
  \frac{1}{g_0} = -\frac{2}{\pi}\int_0^{{\qmax}} \! dp\,p^2
  \frac{[F(p)]^2}{2\Eb{p}}\,,
  \label{gapeqseparable}
\end{equation}
with $\Eb{p} = \sqrt{\xib{p}^2+[\Deltab{0} F(p)]^2}$. Using the
relationship between $g_0$ and the scattering length $a$ in free
space,
\begin{equation}
  \frac{1}{a} = \frac{1}{mg_0}+\frac{2}{\pi}\int_0^{{\qmax}}\! dp\, [F(p)]^2\,,
  \label{renormalization}
\end{equation}
one can rewrite \Eq{gapeqseparable} in the form
\begin{equation}
  \frac{1}{a} = -\frac{2}{\pi}\int_0^{\qmax}\! dp\,[F(p)]^2
  \left(\frac{p^2}{2m\Eb{p}} - 1\right),
  \label{gap-equation-renormalized}
\end{equation}
which has the advantage that the integrand tends toward zero more
smoothly, already before $p$ approaches the cutoff $\qmax$, and one
can see that the gap equation remains well defined in the
limit $\Lambda \to \infty$.

So far, we have only discussed the gap but not the mean field
$\UHF{k}$. In the literature on ultracold atoms, the latter is usually
not taken into account, because it vanishes in the limit
$\Lambda\to\infty$ \cite{Pieri2004}. It is given by
\begin{equation}
  \UHF{k} = \int\! \frac{d^3p}{(2\pi)^3}
  \left(\frac{1}{2}-\frac{\xib{p}}{2\Eb{p}}\right)4\pi V_0\Big(\frac{\pv-\kv}{2},\frac{\pv-\kv}{2}\Big)\,.
  \label{UHFB-difficult}
\end{equation}
For the numerical calculation of $\UHF{k}$, it is useful to define an
interaction that is averaged over the angle $\theta$ between $\kv$ and
$\pv$ as follows:
\begin{equation}
\bar{V}_0(k,p) = \frac{1}{2}\int_{-1}^1\! d\cos\theta\; V_0\Big(\frac{\pv-\kv}{2},\frac{\pv-\kv}{2}\Big)\,.
\end{equation}
In terms of this angle-averaged interaction, the mean field can now be
written as
\begin{equation}
  \UHF{k} = \frac{1}{\pi} \int_0^{\qmax}\! dp\, p^2
  \left(1-\frac{\xib{p}}{\Eb{p}}\right) \bar{V}_0(k,p)\,.
\end{equation}
It turns out that $\UHF{k} = 0$ for $k>3\qmax$.

For a given chemical potential $\mu$, the HFB density reads
\begin{equation}
  n_{\HFB} = \frac{N_{\HFB}}{\calV}=
    \frac{1}{2\pi^2} \int_0^{\qmax}\! dk\,k^2
    \left(1-\frac{\xib{k}}{\Eb{k}}\right).
    \label{number-equation}
\end{equation}
Therefore, if one wants to obtain results for a given density $n$ and
not for a given chemical potential $\mu$, one has to determine $\mu$
by solving the equation $n=n_{\HFB}(\mu)$ simultaneously with the gap
equation.

Finally, the ground-state energy density is given by
\begin{equation}
  \frac{\calE_{\HFB}}{\calV}
  = \frac{1}{4\pi^2} \int_0^{\qmax} \! dk\, k^2
  \left[
    \left(1-\frac{\xib{k}}{\Eb{k}}\right)\left(\frac{k^2}{m}+\UHF{k}\right)
    -\frac{\Deltab{k}^2}{\Eb{k}}\right].
\end{equation}

\section{Bogoliubov many-body perturbation theory}
\label{sec:bmbpt}
\subsection{Perturbative corrections to the HFB ground-state energy}
The HFB theory may be a good starting point, but generally corrections
to it are needed. In particular, the energy shift due to the HF field
that we have just discussed is roughly proportional to the coupling
constant $g_0$ and hence strongly dependent on the choice of the
cutoff $\Lambda$, whereas physical results should be of course cutoff
independent. We therefore expect that, by including higher-order
corrections, if these converge to the exact value of the ground-state
energy, the cutoff dependence should cancel out.

Expressing in $\Kop$ the operators $\adag{}$ and $\aop{}$ in terms of
quasiparticle operators $\bdag{}$ and $\bop{}$, and normal ordering
with respect to these (i.e., putting all $\bdag{}$ operators to the
left of the $\bop{}$ operators), one can write $\Kop$ in the form
\begin{equation}
  \Kop = K_{00}+\Kop_{11}+\Kop_{40}+\Kop_{31}+\Kop_{22}+\Kop_{13}+\Kop_{04}\,,
\end{equation}
where $\Kop_{ij}$ represents the terms having products of $i$
quasiparticle creation operators $\bdag{}$ followed by $j$
quasiparticle annihilation operators $\bop{}$. Notice that there are
no terms $\Kop_{20}$ and $\Kop_{02}$ with two $\bdag{}$ or two
$\bop{}$ operators, because these terms vanish if the $u$ and $v$
factors are determined according to \Eq{uvfactors}. The first term,
$K_{00}$, is just a c-number and obviously it corresponds to the
expectation value of $\Kop$ in the state that has no quasiparticles,
i.e., in the HFB ground state:
\begin{equation}
K_{00} = \calE_{\HFB}-\mu\, N_{\HFB}\,.
\end{equation}

The second term, $\Kop_{11}$, has the simple form
\begin{equation}
\Kop_{11} = \sum_{\kv\sigma} \Eb{k} \bdag{\kv\sigma}\bop{\kv\sigma}\,.
\end{equation}
Hence, the eigenstates and eigenvalues of
\begin{equation}
  \Kop_0 = K_{00}+\Kop_{11}
\end{equation}
are simple and known: its ground state is the HFB state $\ket{0^{(0)}}
= \ket{\HFB}$ with eigenvalue $\Omega_0^{(0)} = K_{00}$, the lowest
excited states are one-quasiparticle (1qp) states
$\ket{(\kv\sigma)^{(0)}} = \bdag{\kv\sigma}\ket{\HFB}$ with
eigenvalues $\Omega^{(0)}_{\kv\sigma} = K_{00}+\Eb{k}$, followed by 
the two-quasiparticle (2qp) states
$\ket{(\kv_1\sigma_1\kv_2\sigma_2)^{(0)}} =
\bdag{\kv_1\sigma_1}\bdag{\kv_2\sigma_2}\ket{\HFB}$ with eigenvalues
$\Omega^{(0)}_{\kv_1\sigma_1\kv_2\sigma_2} = K_{00}+\Eb{k_1}+\Eb{k_2}$,
and so on.

We will denote the remaining terms of $\Kop$ as
\begin{equation}
  \Wop = \Kop_{40}+\Kop_{31}+\Kop_{22}+\Kop_{13}+\Kop_{04}
\end{equation}
and introduce a formal parameter $\lambda$ (where the physical
situation corresponds to $\lambda=1$) to write
\begin{equation}
  \Kop = \Kop_0+\lambda \Wop\,.
  \label{formallambda}
\end{equation}
Following, e.g., the book by Sakurai \cite{Sakurai}, we can now apply
time-independent perturbation theory, i.e., an expansion of the
eigenstates and eigenvalues of $\Kop$ in powers of $\lambda$, where
$\Kop_0$, $\Wop$, and $\Omega^{(0)}_{\mu}$ play the roles of the
unperturbed hamiltonian $H_0$, the perturbation $V$, and the
unperturbed energies $E^{(0)}_\mu$, respectively. This is exactly what
was done in \myRef{Tichai2018} and called there Bogoliubov many-body
perturbation theory (BMBPT), generalizing the usual many-body
  perturbation theory (MBPT) on top of HF to the case with
  pairing. To fix our notations, which are slightly different from
those of \myRef{Tichai2018}, we write the expansion of the ground
state $\ket{0}$ up to some order $n$ and the corresponding eigenvalue
$\Omega_0$ as
\begin{equation}
  \ket{0}\approx \sum_{i=0}^{n} \lambda^i \ket{0^{(i)}}\,,\quad
  \Omega_0\approx \sum_{i=0}^{n+1} \lambda^i \Omega_0^{(i)}\,.
  \label{stateexpansion}
\end{equation}
The leading ($i=0$) contributions correspond to the HFB result.
The first corrections to the ground state are given by
\begin{gather}
  \ket{0^{(1)}} = \sum_{\mu\neq 0} \ket{\mu^{(0)}}
    \frac{\bra{\mu^{(0)}}\Wop\ket{0^{(0)}}}
       {\Omega_0^{(0)}-\Omega_{\mu}^{(0)}}\,,\label{state1st}\\
  \ket{0^{(2)}} = \sum_{\mu\nu\neq 0} \ket{\nu^{(0)}}
  \frac{\bra{\nu^{(0)}}\Wop\ket{\mu^{(0)}}
    \bra{\mu^{(0)}}\Wop\ket{0^{(0)}}}
       {(\Omega_0^{(0)}-\Omega_{\nu}^{(0)})(\Omega_0^{(0)}-\Omega_{\mu}^{(0)})}\,,
       \label{state2nd}
\end{gather}
where we have used the fact that $\Omega_0^{(1)} =
\bra{0^{(0)}}\Wop\ket{0^{(0)}} = 0$. The corresponding corrections to
$\Omega_0$ are
\begin{gather}
  \Omega_0^{(2)} = \sum_{\mu\neq 0}
    \frac{\bra{0^{(0)}}\Wop\ket{\mu^{(0)}}\bra{\mu^{(0)}}\Wop\ket{0^{(0)}}}
       {\Omega_0^{(0)}-\Omega_{\mu}^{(0)}}\,,\label{eigenvalue2nd}\\
  \Omega_0^{(3)} = \sum_{\mu\nu\neq 0}
  \frac{\bra{0^{(0)}}\Wop\ket{\nu^{(0)}}\bra{\nu^{(0)}}\Wop\ket{\mu^{(0)}}
    \bra{\mu^{(0)}}\Wop\ket{0^{(0)}}}
       {(\Omega_0^{(0)}-\Omega_{\nu}^{(0)})(\Omega_0^{(0)}-\Omega_{\mu}^{(0)})}\,.
       \label{eigenvalue3rd}
\end{gather}
At higher orders, there appear ``disconnected diagrams'', which have
to be discarded \cite{ShavittBartlett}, but this does not yet happen
at the second and third orders considered here. In
practice, since the only part of $\Wop$ that gives a non-vanishing
result when acting on $\ket{0^{(0)}}$ is $\Kop_{40}$, the index $\mu$
runs only over 4qp states with two $\up$ and two $\down$
quasiparticles. The same is true for the index $\nu$ in
\Eq{eigenvalue3rd} for $\Omega_0^{(3)}$, where the operator that acts
on $\bra{0^{(0)}}$ to the left must be $\Kop_{04}$. Consequently, in
this equation, the $\Wop$ operator in the middle must not change the
number of quasiparticles, and therefore it must be $\Kop_{22}$. The
situation is more difficult in \Eq{state2nd} for the state
$\ket{0^{(2)}}$, where the sum over $\nu$ has to include also 2qp and
6qp states, which are generated when $\Kop_{13}$ and $\Kop_{31}$ act
on the 4qp state $\ket{\mu^{(0)}}$.

From now on, since we are only interested in the ground state, we will
simply write $\Omega$ instead of $\Omega_0$ for the lowest eigenvalue.

The question arises how one can compute, e.g., the energy $\calE$ as a
function of the density $n=N/\calV$. The problem is that $\Nop$ does
not commute with $\Kop_0$, although it commutes of course with
$\Kop$. Therefore, as it is well known in HFB theory, the eigenstates
of $\Kop_0$ are not eigenstates of $\Nop$, and one may wonder what
value one should use for $N$. Interestingly, one can show that up to
order $\lambda^3$, the energy $\calE$ as a function of the density $n
= N/\calV$ can be immediately obtained from the expansion of the grand
potential $\Omega$ as follows:
\begin{equation}
  \calE(n_{\HFB}(\mu)) = \calE_{\HFB}(\mu) + \lambda^2
  \Omega^{(2)}(\mu) + \lambda^3 \Omega^{(3)}(\mu) +
  \calO(\lambda^4)\,.
  \label{energy_from_omega}
\end{equation}
where $\Omega^{(2)}$ and $\Omega^{(3)}$ are second- and third-order 
corrections. These are explicitly computed in the following two subsections.

\subsection{Second-order BMBPT correction}
\label{sec:sec-order-bmbpt}
\begin{figure}
  \epsfig{width=7.8cm,file=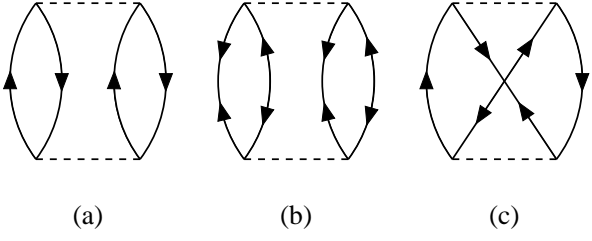}
  \caption{\label{fig:2ndorder} Goldstone-like diagrams corresponding to
    the three terms in the second-order BMBPT contribution to the
    ground-state energy. Lines with upwards pointing arrows represent
    particles (factor $u^2$), lines with downwards pointing arrows
    represent holes (factor $v^2$), and lines with two opposite arrows represent
    anomalous propagators (factor $uv$). The dashed lines represent
    the interaction. Diagram (a) is the only one that exists 
    in the limit of no pairing.}
\end{figure}

As mentioned before, the second-order correction (\ref{eigenvalue2nd})
requires to sum over 4qp states having two $\up$ and two $\down$
quasiparticles and zero total momentum ($\kv_1+\kv_2+\kv_3+\kv_4=0$)
\begin{equation}
  \ket{\mu^{(0)}} =
  \bdag{\kv_1\up}\bdag{\kv_2\down}\bdag{\kv_3\up}\bdag{\kv_4\down}\ket{0^{(0)}}
\end{equation}
with energies
\begin{equation}
  \Omega_{\mu}^{(0)} = \Omega_0^{(0)} + \Eb{k_1}+\Eb{k_2}+\Eb{k_3}+\Eb{k_4}\,.
\end{equation}
Notice that permutations of $\kv_1$ and $\kv_3$ or of $\kv_2$ and
$\kv_4$ do not generate a different state, so when integrating over
all momenta one has to divide by 4. Therefore,
\Eq{eigenvalue2nd} becomes
\begin{equation}
\Omega^{(2)} = -\frac{1}{4}\sum_{\kv_1\kv_2\kv_3}
    \frac{|\bra{0^{(0)}}\Kop_{04}\bdag{\kv_1\up}\bdag{\kv_2\down}\bdag{\kv_3\up}\bdag{\kv_4\down}\ket{0^{(0)}}|^2}
       {\Eb{k_1}+\Eb{k_2}+\Eb{k_3}+\Eb{k_4}}\,,\label{Omega2nd}\\
\end{equation}
with $\kv_4 = -\kv_1-\kv_2-\kv_3$\,. The explicit form of $\Kop_{04}$
is obtained by inserting \Eq{bogoliubovtrafo} into \Eq{hamiltonian}
with \Eq{Vpartialwaves} and keeping only the term with four $\bop{}$
operators
\begin{multline}
  \Kop_{04} = -\frac{4\pi}{\calV}\sum_{\pv_1\dots\pv_4}
  V_0\big(q_{12},q_{34}\big)\delta_{\pv_1+\pv_2+\pv_3+\pv_4,\vek{0}}\\
  \times u_{p_1}u_{p_2}v_{p_3}v_{p_4}
  \bop{\pv_4\down}\bop{\pv_3\up}\bop{\pv_2\down}\bop{\pv_1\up}
\end{multline}
where $q_{ij} = |\kv_i-\kv_j|/2$. For convenience, we have renamed the
original momentum labels $\pv_1, \pv_2, \pv'_1, \pv'_2$ of
\Eq{hamiltonian} into $-\pv_4, -\pv_3, \pv_1, \pv_2$ and we have used
$V_0(q,q')=V_0(q',q)$. It is straight-forward to work out the
  matrix element in the numerator of \Eq{Omega2nd}:
\begin{multline}
  \bra{0^{(0)}}\Kop_{04}\bdag{\kv_1\up}\bdag{\kv_2\down}\bdag{\kv_3\up}\bdag{\kv_4\down}\ket{0^{(0)}} = \\
  -\frac{4\pi}{\calV}
    [V_0(q_{12},q_{34})(u_1u_2v_3v_4+v_1v_2u_3u_4)\\
      -V_0(q_{14},q_{32})(v_1u_2u_3v_4+u_1v_2v_3u_4)]\,,
\end{multline}
where $u_{k_i}$ and $v_{k_i}$ are denoted as $u_i$ and $v_i$ for 
better readability. The square
of this expression gives ten terms, which after a suitable relabeling
of the momenta (leaving the denominator of \Eq{Omega2nd} unchanged)
can be grouped together into three terms. The final expression reads
\begin{gather}
  \Omega^{(2)} = -\frac{(4\pi)^2}{\calV^2} \sum_{\kv_1\kv_2\kv_3}
  \frac{A+B+C}
       {\Eb{1}+\Eb{2}+\Eb{3}+\Eb{4}}\,,\\
  A = v_1^2\, v_2^2\,u_3^2\,u_4^2\,
       [V_0(q_{12},q_{34})]^2\,,\\
  B = u_1v_1\, u_2v_2\, u_3v_3\, u_4v_4\,
       [V_0(q_{12},q_{34})]^2\,,\\
  C = -2\,u_1v_1\,v_2^2\,u_3v_3\,u_4^2\,
       V_0(q_{12},q_{34})\,V_0(q_{14},q_{23})\,,
\end{gather}
where $\Eb{i} = \Eb{k_i}$. These three terms can be interpreted
  diagrammatically as the three Goldstone-like diagrams shown in
  \Fig{fig:2ndorder}.

Notice that, in the limit of no pairing ($\Deltab{k}=0$), where $u_k^2
= \theta(k-\kF)$, $v_k^2 = \theta(\kF-k)$, and $u_kv_k = 0$, only the
term $A$ contributes while the terms $B$ and $C$ vanish. Then the
energy denominator becomes $\xi_3+\xi_4-\xi_1-\xi_2$, and we recover
the usual second-order correction to the HF energy.

In practice, the sums over the $\kv_i$ are replaced by integrals
according to \Eq{summegral}, so that $\Omega^{(2)}$ becomes
proportional to the volume $\calV$ as it should be, since $\Omega$ is
related to the pressure $P$ by $\Omega = -P\calV$.

The integrations are done with Monte-Carlo sampling.  Notice that for
each integration variable $\kv_i$, the integral can be written in the
form $\int\! d^3k_i\, w(k_i)\, f(k_i,\theta_i,\phi_i)$, where $w(k)$ is
one of the functions $v_k^2$, $u_kv_k$, or $u_k^2$, and all the angle
dependence is in the remaining function $f$. If the integrand contains
a factor of $v_k^2$ or $u_kv_k$, the integration region is
automatically limited to $k<\qmax$, because $v_k=0$ for
$k>\qmax$. However, if the integrand contains a factor of $u_k^2$, as
it is the case for the $\kv_3$ integration in term $A$, it is only cut
off through the interaction term $[V_0(q_{12},q_{34})]^2$ contained in
the function $f$, which vanishes for $q_{34}>\qmax$. In this case,
combining the constraints $k_1, k_2, q_{34} < \qmax$ and momentum
conservation $\kv_3 = (2\qv_{34}-\kv_1-\kv_2)/2$, one sees that $k_3$
is limited to the region $k_3 < 2\qmax$. We implement importance
sampling to efficiently distribute the integration points along $k_i$,
according to the weights $k_i^2 w(k_i)$ by introducing three
transformations of variables. For each function $w(k)$, we define a
function
\begin{equation}
  x_w(p) = \int_0^p\! dk\,k^2\, w(k)\,.
\end{equation}
We also define the corresponding inverse functions $p_w(x)$ such that
$p_w(x_w(p)) = p$. Hence, we can write
\begin{multline}
  \int\! d^3k_i\, w(k_i)\, f(k_i,\theta_i,\phi_i)\\
  = \int_0^{x_w(k_{i\max})}\!dx_i
  \int_{-1}^1\! d\cos\theta_i \int_0^{2\pi}\! d\phi_i\,
  f(p_w(x_i),\theta_i,\phi_i)\,,\label{importance-sampling}
\end{multline}
with $k_{i\max} = \qmax$ or $2\qmax$, respectively, as discussed
above. The advantage of these transformations is that now uniformly
distributed random variables $x_i$ correspond to momenta
$k_i=p_w(x_i)$ whose distributions automatically account for the
factors $k_i^2 w(k_i)$ in the integrand.

Finally, because of rotational invariance, the integrand depends only
on relative angles. Therefore, we may choose without any loss of
generality $\kv_1$ in $z$ direction and $\kv_2$ in the $xz$ plane, so
that the integrations over $\cos \theta_1$, $\phi_1$, and $\phi_2$
become trivial.
\subsection{Third-order BMBPT correction}
\label{sec:thirdorder}
\begin{figure}
  \epsfig{width=7.8cm,file=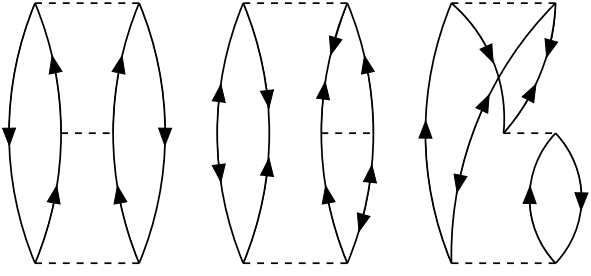}
  \caption{\label{fig:3rdorder} Three examples out of the 27 distinct
    Goldstone-like diagrams for the third-order BMBPT correction to
    the ground-state energy.}
\end{figure}

According to \Eq{eigenvalue3rd} and the discussion below that
equation, the third-order correction is given by
\begin{multline}
\Omega_0^{(3)} = \frac{1}{16}\sum_{\kv_1\dots \kv_8}
  \frac{1}{\Eb{1,2,3,4}\Eb{5,6,7,8}}\\
\times \bra{0^{(0)}}\Kop_{04}
  \bdag{\kv_1\up}\bdag{\kv_2\down}\bdag{\kv_3\up}\bdag{\kv_4\down}\ket{0^{(0)}}\\
\times
  \bra{0^{(0)}}\bop{\kv_4\down}\bop{\kv_3\up}\bop{\kv_2\down}\bop{\kv_1\up}
  \Kop_{22}
  \bdag{\kv_5\up}\bdag{\kv_6\down}\bdag{\kv_7\up}\bdag{\kv_8\down}\ket{0^{(0)}}\\
\times
  \bra{0^{(0)}}\bop{\kv_8\down}\bop{\kv_7\up}\bop{\kv_6\down}\bop{\kv_5\up}
  \Kop_{40}\ket{0^{(0)}}\,,
\label{thirdorder4qp}
\end{multline}
with the abbreviations
\begin{equation}
\Eb{1,2,3,4} = \Eb{k_1}+\Eb{k_2}+\Eb{k_3}+\Eb{k_4}
\end{equation}
and so on. Analogous to the factor $1/4$ in \Eq{Omega2nd}, the
factor $1/16$ in \Eq{thirdorder4qp} takes into account that
permutations $\kv_1\leftrightarrow \kv_3$, $\kv_2\leftrightarrow
\kv_4$, etc., describe the same state and therefore must be counted
only once. Momentum conservation in $\Kop_{04}$ and $\Kop_{40}$
require that $\kv_1+\kv_2+\kv_3+\kv_4 = \kv_5+\kv_6+\kv_7+\kv_8 =
0$. The explicit expression of $\Kop_{22}$ reads
\begin{multline}
  \Kop_{22} = \frac{4\pi}{\calV}\sum_{\pv_1\dots\pv_4}
  \delta_{\pv_1+\pv_2,\pv_3+\pv_4}\\
  \times\big\{\big[V_0\big(\tfrac{|\pv_1-\pv_2|}{2},\tfrac{|\pv_3-\pv_4|}{2}\big)
      (u_{p_1}u_{p_2}u_{p_3}u_{p_4}+v_{p_1}v_{p_2}v_{p_3}v_{p_4})\\
    +V_0\big(\tfrac{|\pv_1+\pv_3|}{2},\tfrac{|\pv_2+\pv_4|}{2}\big)
    (u_{p_1}v_{p_2}v_{p_3}u_{p_4}+v_{p_1}u_{p_2}u_{p_3}v_{p_4})\big]\\
    \times\bdag{\pv_1\up}\bdag{\pv_2\down}\bop{\pv_4\down}\bop{\pv_3\up}\\
    -V_0\big(\tfrac{|\pv_1+\pv_4|}{2},\tfrac{|\pv_2+\pv_3|}{2}\big)
    \big[u_{p_1}v_{p_2}u_{p_3}v_{p_4}
      \bdag{\pv_1\up}\bdag{\pv_2\up}\bop{\pv_4\up}\bop{\pv_3\up}\\
      +v_{p_1}u_{p_2}v_{p_3}u_{p_4}
      \bdag{\pv_1\down}\bdag{\pv_2\down}\bop{\pv_4\down}\bop{\pv_3\down}
      \big]\big\}\,.
\end{multline}
One can easily see that $\Kop_{22}$ changes only two out of the four
quasiparticle momenta in \Eq{thirdorder4qp}. So, finally, there are
only four independent momenta over which we have to integrate. Like
the second-order correction, the third-order correction can again be
interpreted in terms of Goldstone-like diagrams. Some examples for
such diagrams are shown in \Fig{fig:3rdorder}.
By relabeling the momenta and combining terms with the same weight
functions, one obtains finally an expression which is suitable for
numerical integration using the importance-sampling method given in
\Eq{importance-sampling}. The explicit formula is given in Appendix
\ref{app:Omega3}.

As it was the case for the second-order correction, the third-order
correlation energy without pairing can be obtained by setting in this
expression $\Delta_k = 0$, $v_k^2 = \theta(\kF-k)$, $u_k^2 =
\theta(k-\kF)$, and $u_k v_k = 0$.
\section{Results}
\label{sec:results}
For a given interaction, the equation of state is given by the energy
density $\calE/\calV$ as a function of the density $n = N/\calV$. In
the case of a contact interaction, which is determined by the
scattering length $a$, it can be reduced to a dimensionless function
$\calE/\calE_0$, depending on one dimensionless parameter $1/(\kF a)$,
where $\kF = (3\pi^2 n)^{1/3}$ is the Fermi momentum and
$\calE_0/\calV = \kF^5/(10\pi^2 m)$ is the energy density of the ideal
Fermi gas.

In our case, a complication arises from the cutoff $\Lambda$, leading
to an additional dependence on the dimensionless parameter
$\Lambda/k_F$. Ideally, the results should be independent of this
unphysical parameter, but of course the cutoff must be always large
enough to include all relevant degrees of freedom, i.e., at least
$\Lambda > \kF$.
\begin{figure*}
  \epsfig{width=17.5cm,file=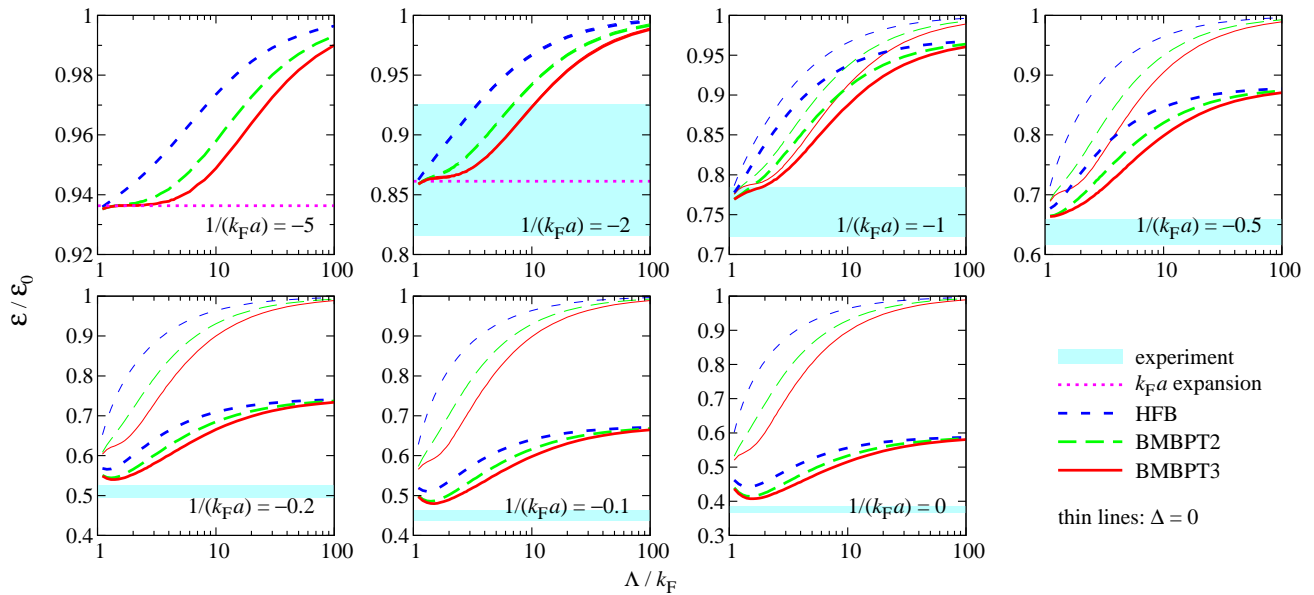}
  \caption{\label{fig:E-Lambda} Cutoff dependence of the computed
    ground-state energy $\calE$ in units of the energy of an ideal
    Fermi gas, $\calE_0$, for seven values of the interaction
    parameter $1/(\kF a)$ at different levels of approximation. Thick
    lines start from the HFB ground state: $\calE_{\HFB}$ (blue short
    dashes), $\calE_{\BMBPT2} = \calE_{\HFB} + \Omega^{(2)}$ (green
    long dashes), $\calE_{\BMBPT3} = \calE_{\HFB} + \Omega^{(2)} +
    \Omega^{(3)}$ (red solid line), while the corresponding thin lines
    are obtained without pairing, i.e., starting from HF instead of
    HFB. For comparison, the light blue areas are the experimental
    results of \myRef{Horikoshi2017} and the purple dots in the first
    two panels are obtained from the $\kF a$ expansion up to order
    $(\kF a)^4$ of \myRef{Wellenhofer2021}.}
\end{figure*}
\Fig{fig:E-Lambda} shows the cutoff dependence of the ground-state
energy $\calE$ in units of $\calE_0$, obtained according
to \Eq{energy_from_omega}, for different values of the parameter
$1/(\kF a)$. The thick lines are the HFB (+ BMBPT)
results, while the thin lines are HF (+ MBPT) results.

Let us start our discussion with the first two panels, $1/(k_Fa) = -5$
and $-2$, corresponding to the weak-coupling regime. In this regime,
the pairing gap is so small that the thick and thin lines practically
coincide, and the dominant contribution comes from the HF
self-energy. Since the HF contribution is proportional to the coupling
constant $g_0$, which tends toward zero for $\Lambda\to\infty$, it is
not surprising that the HF(B) result (short blue dashes) shows a
strong cutoff dependence. However, the situation improves once the
perturbative corrections are included (green long dashes and red solid
lines). In some range of not too large cutoffs ($\Lambda/\kF \lesssim
2.5$), the cutoff dependence of the HF(B) energy is
compensated by the cutoff dependence of the perturbative corrections
already at third order. Furthermore, in this range of cutoffs, the
perturbation expansion of the energy converges to the energy obtained
in \myRef{Wellenhofer2021} from an expansion in powers of $\kF a$ up
to fourth order (purple dots).

When the interation strength increases, the pairing gaps get bigger.
Therefore, in the range $-1 \leqslant 1/(\kF a) \leqslant 0$, where
the gas is strongly correlated, the HFB + BMBPT results markedly
differ from the HF + MBPT ones. In particular, only when starting
  from HFB, one obtains a finite correction to the energy in the limit
  $\Lambda\to\infty$. As $1/(\kF a)$ approaches $0$, which is the
unitary limit, the results are always cutoff dependent. However, it is
worth nothing that with the HFB + BMBPT, the cutoff dependence between
the different orders of the perturbation theory is less compared to
the HF + MBPT results. In the HFB + BMBPT case, we observe that,
  as a function of the cutoff, the energy has a minimum around
  $\Lambda/\kF \sim 1.5$. 

Various thermodynamic quantities for the homogenous one-component
Fermi gas from the BCS regime to the unitary limit were recently
determined in~\cite{Horikoshi2017} using $^6$Li atoms in a hybrid
trap. The results for the energy with their error bands are shown in
Fig.~\ref{fig:E-Lambda} by the light blue regions. Our minimum
energies agree rather well with these experimental results. Although
we do not see true convergence as in the weakly interacting case, one
may argue that these minimum energies can be considered our best
estimates because the third-order change is much smaller than the
second-order one, and also there is at least local cutoff
independence. However, as one approaches the unitary limit, the best
estimate from HFB + BMBPT is still somewhat higher than the
experimentally determined value of the energy. The ratio of
$\calE/\calE_0$ in the unitary limit is called the Bertsch parameter
$\xi$. At the respective minima, we find $\xi_{\HFB} = 0.442$,
$\xi_{\BMBPT2} = 0.414$, and $\xi_{\BMBPT3} = 0.407$, while the most
precise experimental value is $\xi = 0.370 \pm 0.005$ \cite{Ku2012},
which lies inside the error band of \cite{Horikoshi2017} shown in our
figure and agrees also very well with the Quantum Monte Carlo results
$\xi=0.372 \pm 0.005$ \cite{Carlson2011} and $\xi=
0.366^{+0.016}_{-0.011}$ \cite{Endres2013}. Comparing the HFB and
BMBPT energies, we observe that the minima get broader at higher
orders of perturbation theory. This might indicate the onset of
convergence, which should eventually lead to cutoff independence at
least in some range of cutoffs, as discussed above for the weakly
interacting case.

\section{Conclusion and open questions}
\label{sec:concl}
Inspired by the $\vlowk$ interactions used in nuclear physics, we have
constructed effective interactions which tend to zero above some
momentum cutoff $\Lambda$ but preserve exactly the scattering phase
shifts of the contact interaction below that cutoff. Since the phase
shifts do not change sign, one can obtain rank-one separable
interactions with these properties using the inverse-scattering
formula of Tabakin~\cite{Tabakin1969}, without solving explicitly the
RG evolution equation.

Using these separable interactions, we have calculated the equation of
state of the zero-temperature Fermi gas within the HFB + BMBPT
approach. In the weak-coupling BCS regime, where the HF term is much
more important than pairing, we find that, for $\Lambda/k_F \lesssim
2.5$, the BMBPT converges quickly to the correct result. At stronger
coupling, we do not yet find convergence but nevertheless the results
are encouraging and our best estimate for the Bertsch parameter in the
unitary limit, $\xi = 0.407$, is not very far from the experimental
value $\xi = 0.370 \pm 0.005$ \cite{Ku2012}.

An obvious problem of the approach is the cutoff dependence of the
results. For observables that are insensitive to the short-range
scales, the fact that the interaction gives by construction cutoff
independent results in the two-body sector (phase shifts) at low
momentum implies that any cutoff dependence in the many-body sector is
indicative of missing contributions. These missing contributions can
be higher-order corrections in the perturbative expansion or missing
three- and higher-body interactions.

Even if the simplest three-body interaction of the form
$(\psi^\dagger\psi)^3$ is absent in the limit $\Lambda\to\infty$
because it is forbidden by the Pauli principle for a two-component
Fermi system, more complicated terms involving derivatives, i.e.,
momenta, would be generated in the RG evolution when the cutoff is
lowered \cite{Tolos2008,Hebeler2010,Hammer2013}. These terms
should be either included explicitly or, in an approximate way, in the
form of a density-dependent two-body interaction.

Concerning the perturbation expansion in BMBPT, it is well known from
nuclear structure theory that this can only work if the interactions
are soft enough. This means that, while the cutoff must be larger than
$\kF$ to describe the interactions among the particles in the Fermi
sea, it should not be too large compared to $\kF$. If the cutoff is
too large, e.g., in the limit $\Lambda\to\infty$, non-perturbative
resummations (ladder diagrams) are necessary even in the weakly
coupled regime.  In order to see in which range of $\Lambda/\kF$ and
$1/(\kF a)$ the BMBPT expansion converges, it would be desirable to
push it to higher orders.  This necessitates, however, the development
of tools that can compute these corrections automatically, as it was
done for nuclear-structure theory \cite{Arthuis2019}. Once the
equations have been derived, their numerical computation using
Monte-Carlo integration would not be any more difficult than the third
order that we have done here. As usual with Monte-Carlo integration,
the computation time depends on the precision that one asks for. Let
us mention that a completely different strategy to sum the
perturbative corrections to much higher orders is the diagrammatic
Monte-Carlo method \cite{Spada2021}.

In the strongly coupled regime, it may be necessary to resum
ladder-like diagrams even in the case of small cutoffs, for the
following reason. The BMBPT expansion as explained in \Sec{sec:bmbpt}
is based on excited states built out of fermionic quasiparticle
excitations. But on the BEC ($a>0$) side of the crossover, it is clear
that the most relevant excitations are bosonic ones. Therefore it is
possible that bosonic excitations, namely, the Bogoliubov-Anderson
(BA) phonon, play also some role at unitarity and for $a<0$
\cite{Pieri2004}. This collective mode is described in the superfluid
version of the random-phase approximation (RPA) \cite{Combescot2006}
(called quasiparticle RPA in the nuclear-physics literature),
corresponding to particle-particle and particle-hole ladder diagrams
which are coupled to one another due to the anomalous propagators. For
a regularized contact interaction in the limit $\Lambda\to\infty$, the
effect of the BA mode on the ground state was included in
\myRef{Pieri2004}. Recently it was also studied in \myRef{Inotani2020}
for the case of dilute neutron matter with a separable interaction,
however, without the HF field.

In such T-matrix approaches, the use of a finite cutoff would have
some advantages compared to the limit $\Lambda\to\infty$. At weak
coupling, one gets the right result almost for free (at the HF level)
with low-momentum interactions whereas for $\Lambda\to\infty$ one has
to use at least partially self-consistent versions of the T-matrix
theory \cite{Pini2019} such as the extended T-matrix approximation
\cite{Kashimura2012} or fully self-consistent Green's functions
\cite{Haussmann1994}. In addition, low-momentum interactions may
simplify the numerical implementation of these approaches because a
smaller grid in momentum space is required. The price to pay are
uncertainties due to the unknown many-body interactions that are in
principle induced when one lowers the cutoff.

Furthermore, one might wonder whether the HFB ground state is the best
starting point for the perturbative expansion, although it is known
that the gap is reduced due to ``screening'' of the interaction by the
surrounding medium \cite{Gorkov1961}. We leave all these open
questions for future work.
\begin{acknowledgments} We thank T. Duguet for discussions and
    H. Tajima for sending us the data of \myRef{Horikoshi2017}. We
    acknowledge support from the Collaborative Research Program of
    IFCPAR/CEFIPRA, Project number: 6304-4.
  \end{acknowledgments}

\appendix
\section{Details of the computation of the separable interaction}
Let us first consider the case $a<0$. When solving the inverse
scattering problem, \Eqs{inversescattering} and \eqref{g0F}, it is
helpful to determine the coupling constant $g_0$ by considering
\Eq{inversescattering} in the limit $q\to 0$:
\begin{equation}
  g_0 = \frac{a}{m} \exp\left(-\frac{2}{\pi}\int_0^{\qmax}\! dq'\,
  \frac{\delta(q')}{q'}\right).
  \label{defg0}
\end{equation}
Then the form factor is given by
\begin{equation}
  F(q) = \sqrt{\frac{V_0(q,q)}{g_0}} = \sqrt{\frac{\sin\delta(q)}{-qa}}
  \exp\left(-\frac{x(q)}{2\pi}\right),
\end{equation}
with
\begin{equation}
  x(q) = 2q \int_0^{\qmax}\! dq'\,
  \frac{q\delta(q')-q'\delta(q)}
       {q'(q^{\prime\,2}-q^2)}
       +\delta(q)\ln\frac{\qmax-q}{\qmax+q}
       \label{defxq}
\end{equation}
for $0<q<\qmax$. Furthermore, $F(0) = 1$ and we set $F(q \geqslant
\qmax) = 0$ since it is negligible by the definition of $\qmax$.

Let us now consider the unitary limit, $a\to\infty$. In this case, the
coupling constant can be written as
\begin{equation}
  g_0 = -\frac{1}{m\qmax} \exp\left(-\frac{2}{\pi}\int_0^{\qmax}\! dq'
    \frac{\tilde{\delta}(q')}{q'}\right),
\end{equation}
with
\begin{equation}
  \tilde{\delta}(q') = \delta(q')-\delta(0)\,,
\end{equation}
and the form factor for $0<q<\qmax$ is given by
\begin{equation}
  F(q) =\sqrt{\frac{\sin\delta(q)}{\sqrt{1-q^2/\qmax^2}}}
  \exp\left(-\frac{\tilde{x}(q)}{2\pi}\right),
\end{equation}
where $\tilde{x}(q)$ is defined analogously to \Eq{defxq} with
$\delta(q')$ replaced by $\tilde{\delta}(q') = \delta(q')-\pi/2$.

Finally, let us consider the case $a>0$. Then the potential has a
bound state with binding energy $1/(ma^2)$ and the phase shift fulfils
$\delta(0) = \pi$. In this case, the coupling constant can be written as
\begin{equation}
g_0 = -\frac{1}{m\qmax^2 a} \exp\left(-\frac{2}{\pi}\int_0^{\qmax}\! dq'
    \frac{\tilde{\delta}(q')}{q'}\right),
\end{equation}
and the form factor for $0<q<\qmax$ is given by
\begin{equation}
  F(q) =\sqrt{\frac{1+q^2 a^2}{1-q^2/\qmax^2}\frac{\sin\delta(q)}{qa}}
  \exp\left(-\frac{\tilde{x}(q)}{2\pi}\right),
\end{equation}
where $\tilde{x}(q)$ is defined analogously to \Eq{defxq} with
$\delta(q')$ replaced by $\tilde{\delta}(q') = \delta(q')-\pi$.

\section{Explicit expression of the third-order BMBPT correction}
\label{app:Omega3}
As explained in \Sec{sec:thirdorder}, in each term of $\Omega^{(3)}$,
we must integrate over four independent momentum vectors. We
relabel in each term the indices in such a way that the independent
integration variables are called $\kv_1\dots\kv_4$. The remaining four
momentum vectors are then given by various combinations of these
integration variables, and we denote these combinations as
\begin{gather}
  \kv_5 = -\kv_1-\kv_2-\kv_3\,,\qquad
  \kv_6 = -\kv_1-\kv_2-\kv_4\,,\nonumber\\
  \kv_7 = -\kv_1-\kv_3-\kv_4\,,\qquad
  \kv_8 = -\kv_2-\kv_3-\kv_4\,,\nonumber\\
  \kv_9 = \kv_1+\kv_2-\kv_3\,,\qquad
  \kv_{10} = \kv_1+\kv_3-\kv_2\,,\nonumber\\
  \kv_{11} = \kv_1+\kv_2-\kv_4\,,\qquad
  \kv_{12} = \kv_1+\kv_4-\kv_2\,,\nonumber\\
  \kv_{13} = \kv_2+\kv_3-\kv_4\,.
\end{gather}
The interaction potential appears with various differences or sums of
momenta and we define the notation
\begin{equation}
  F^{\pm}_{i,j} = F(|\kv_i\pm \kv_j|/2)\,.
\end{equation}
Combining terms having the same weight functions, the third-order
correction can be finally written as
\begin{multline}
  \Omega^{(3)} = \frac{(4\pi g_0)^3}{\calV^3}\sum_{\kv_1\dots\kv_4}
  (u_1v_1\, u_2v_2\, u_3v_3\, u_4v_4\, A_1\\
  + u_1v_1\, u_2v_2\, u_3v_3\, v_4^2\, A_2
  + u_1v_1\, u_2v_2\, u_3v_3\, u_4^2\, A_3\\
  + u_1v_1\, u_2v_2\, v_3^2\, v_4^2\, A_4 
  + u_1v_1\, u_2v_2\, v_3^2\, u_4^2\, A_5\\
  + v_1^2\, v_2^2\, v_3^2\, u_4^2\, A_6
  + v_1^2\, v_2^2\, u_3^2\, u_4^2\, A_7)\,,
\end{multline}
with
\begin{multline}
  A_1 =
    -\frac{4 (u_{13}^2 u_5^2 + v_{13}^2 v_5^2) F^-_{2,3} F^-_{4,13}
        F^-_{2,3}F^-_{1,5} F^-_{1,4} F^-_{5,13}}
      {\Eb{1,13,4,5} \Eb{1,2,3,5}}\\
    +\frac{4 u_5v_5 u_6v_6 F^-_{1,3} F^-_{2,5} F^-_{1,4} F^-_{2,6} F^+_{3,4} F^+_{5,6}}
      {\Eb{1,2,3,5} \Eb{1,2,4,6}}\\
    +\frac{2 u_5^2 u_6^2 F^-_{2,3} F^-_{1,5} F^-_{1,4} F^-_{2,6} F^-_{3,5} F^-_{4,6}}
      {\Eb{1,2,3,5} \Eb{1,2,4,6}}\\
    -\frac{4 (u_5^2 v_{13}^2 + u_{13}^2 v_5^2) F^-_{1,3} F^-_{2,5}
        F^+_{3,13} F^+_{2,4} F^-_{4,13} F^-_{1,5}}
      {\Eb{1,13,4,5} \Eb{1,2,3,5}}\\
    +\frac{4 u_6^2 v_5^2 F^-_{1,2} F^-_{3,5} F^-_{1,2} F^-_{4,6} F^+_{3,4} F^+_{5,6}}
      {\Eb{1,2,3,5} \Eb{1,2,4,6}}\\
    +\frac{2 v_5^2 v_6^2 F^-_{1,3} F^-_{2,5} F^-_{2,4} F^-_{1,6} F^-_{3,5} F^-_{4,6}}
      {\Eb{1,2,3,5} \Eb{1,2,4,6}}\,,
\end{multline}
\begin{multline}
  A_2 =
    -\frac{2 u_6^2 u_9v_9 F^+_{1,3} F^+_{2,9} F^-_{2,4} F^-_{1,6} F^-_{3,4} F^-_{6,9}}
      {\Eb{1,2,4,6} \Eb{3,4,6,9}}\\
    -\frac{4 u_6^2 u_5v_5 F^-_{1,3} F^-_{2,5} F^-_{2,4} F^-_{1,6} F^+_{3,4} F^+_{5,6}}
      {\Eb{1,2,3,5} \Eb{1,2,4,6}}\\
    +\frac{2 u_5v_5 v_6^2 F^-_{1,2} F^-_{3,5} F^-_{1,2} F^-_{4,6} F^-_{3,5} F^-_{4,6}}
      {\Eb{1,2,3,5} \Eb{1,2,4,6}}\\
    +\frac{u_9v_9 v_6^2 F^-_{1,2} F^-_{4,6} F^-_{1,2} F^-_{3,9} F^-_{4,6} F^-_{3,9}}
      {\Eb{1,2,4,6} \Eb{3,4,6,9}}\,,
\end{multline}
\begin{multline}
  A_3 =
    \frac{2 u_6^2 u_5v_5 F^-_{1,2} F^-_{3,5} F^-_{1,2} F^-_{4,6} F^-_{3,5} F^-_{4,6}}
      {\Eb{1,2,3,5} \Eb{1,2,4,6}}\\
    +\frac{u_6^2 u_9v_9 F^-_{1,2} F^-_{4,6} F^-_{1,2} F^-_{3,9} F^-_{4,6} F^-_{3,9}}
      {\Eb{1,2,4,6} \Eb{3,4,6,9}}\,,
\end{multline}
\begin{multline}
  A_4 =
    -\frac{4 u_{13}^2 u_5^2 F^-_{2,3} F^-_{1,5} F^+_{3,13} F^+_{2,4} F^-_{1,4} F^-_{5,13}}
      {\Eb{1,13,4,5} \Eb{1,2,3,5}}\\
    +\frac{8 u_{11}^2 u_5^2 F^+_{2,11} F^+_{1,4} F^-_{1,3} F^-_{2,5} F^-_{3,4} F^-_{5,11}}
      {\Eb{1,2,3,5} \Eb{11,3,4,5}}\\
    +\frac{4 u_5^2 u_6^2 F^-_{2,3} F^-_{1,5} F^-_{1,4} F^-_{2,6} F^+_{4,5} F^+_{3,6}}
      {\Eb{1,2,3,5} \Eb{1,2,4,6}}\\
    -\frac{4 u_5^2 u_7^2 F^-_{1,3} F^-_{2,5} F^-_{3,4} F^-_{1,7} F^+_{4,5} F^+_{2,7}}
      {\Eb{1,2,3,5} \Eb{1,3,4,7}}\\
    +\frac{2 u_7^2 u_8^2 F^-_{3,4} F^-_{1,7} F^-_{3,4} F^-_{2,8} F^+_{2,7} F^+_{1,8}}
      {\Eb{1,3,4,7} \Eb{2,3,4,8}}\\
    +\frac{2 u_7^2 v_{12}^2 F^-_{2,12} F^-_{1,4} F^-_{3,12} F^-_{2,7} F^-_{3,4} F^-_{1,7}}
      {\Eb{1,3,4,7} \Eb{12,2,3,7}}\\
    -\frac{4 u_5^2 v_{13}^2 F^-_{1,3} F^-_{2,5} F^-_{2,3} F^-_{4,13} F^-_{4,13} F^-_{1,5}}
      {\Eb{1,13,4,5} \Eb{1,2,3,5}}\,,
\end{multline}
\begin{multline}
  A_5 =
    -\frac{4 u_5^2 u_7^2 F^-_{1,3} F^-_{4,7} F^-_{2,3} F^-_{1,5} F^-_{2,5} F^-_{4,7}}
       {\Eb{1,2,3,5} \Eb{1,3,4,7}}\\
    +\frac{2 u_7^2 u_8^2 F^-_{1,3} F^-_{4,7} F^-_{2,3} F^-_{4,8} F^-_{1,7} F^-_{2,8}}
       {\Eb{1,3,4,7} \Eb{2,3,4,8}}\\
    +\frac{2 u_7^2 v_{10}^2 F^+_{1,10} F^+_{2,3} F^-_{2,10} F^-_{4,7} F^-_{1,3} F^-_{4,7}}
       {\Eb{1,3,4,7} \Eb{10,2,4,7}}\,,
\end{multline}
\begin{multline}
  A_6 =
    -\frac{2 u_6^2 u_7^2 F^-_{1,2} F^-_{4,6} F^-_{1,3} F^-_{4,7} F^+_{3,6} F^+_{2,7}}
       {\Eb{1,2,4,6} \Eb{1,3,4,7}}\\
    +\frac{u_6^2 v_9^2 F^-_{1,2} F^-_{4,6} F^-_{1,2} F^-_{3,9} F^-_{4,6} F^-_{3,9}}
       {\Eb{1,2,4,6} \Eb{3,4,6,9}}\,,
\end{multline}
\begin{equation}
  A_7 = \frac{u_5^2 u_6^2 F^-_{1,2} F^-_{3,5} F^-_{1,2} F^-_{4,6} F^-_{3,5} F^-_{4,6}}
       {\Eb{1,2,3,5} \Eb{1,2,4,6}}\,.
\end{equation}


\end{document}